\documentclass[conference]{IEEEtran}
\IEEEoverridecommandlockouts
\usepackage{cite}
\usepackage{amsmath,amssymb,amsfonts}
\usepackage{algorithmic}
\usepackage{graphicx}
\usepackage{textcomp}
\usepackage{subcaption}
\usepackage{booktabs}
\usepackage{xcolor}
\usepackage{url}

\def\BibTeX{{\rm B\kern-.05em{\sc i\kern-.025em b}\kern-.08em
    T\kern-.1667em\lower.7ex\hbox{E}\kern-.125emX}}
\begin{document}

\title{Retrieval-Augmented Few-Shot Prompting Versus Fine-Tuning for Code Vulnerability Detection\\
}

\author{\IEEEauthorblockN{Fouad Trad}
\IEEEauthorblockA{\textit{Electrical and Computer Engineering} \\
\textit{American University of Beirut}\\
Beirut, Lebanon \\
fat10@mail.aub.edu}
\and
\IEEEauthorblockN{Ali Chehab}
\IEEEauthorblockA{\textit{Electrical and Computer Engineering} \\
\textit{American University of Beirut}\\
Beirut, Lebanon \\
chehab@aub.edu.lb}
}

\maketitle

\begin{abstract}
Few-shot prompting has emerged as a practical alternative to fine-tuning for leveraging the capabilities of large language models (LLMs) in specialized tasks. However, its effectiveness depends heavily on the selection and quality of in-context examples, particularly in complex domains. In this work, we examine retrieval-augmented prompting as a strategy to improve few-shot performance in code vulnerability detection, where the goal is to identify one or more security-relevant weaknesses present in a given code snippet from a predefined set of vulnerability categories. We perform a systematic evaluation using the Gemini-1.5-Flash model across three approaches: (1) standard few-shot prompting with randomly selected examples, (2) retrieval-augmented prompting using semantically similar examples, and (3) retrieval-based labeling, which assigns labels based on retrieved examples without model inference. Our results show that retrieval-augmented prompting consistently outperforms the other prompting strategies. At 20 shots, it achieves an F1 score of 74.05\% and a partial match accuracy of 83.90\%. We further compare this approach against zero-shot prompting and several fine-tuned models, including Gemini-1.5-Flash and smaller open-source models such as DistilBERT, DistilGPT2, and CodeBERT. Retrieval-augmented prompting outperforms both zero-shot (F1 score: 36.35\%, partial match accuracy: 20.30\%) and fine-tuned Gemini (F1 score: 59.31\%, partial match accuracy: 53.10\%), while avoiding the training time and cost associated with model fine-tuning. On the other hand, fine-tuning CodeBERT yields higher performance (F1 score: 91.22\%, partial match accuracy: 91.30\%) but requires additional training, maintenance effort, and resources. These results underscore the value of semantically relevant example selection for few-shot prompting and position retrieval-augmented prompting as a practical trade-off between performance and deployment cost in code analysis tasks.
\end{abstract}

\begin{IEEEkeywords}
Retrieval-Augmented Generation, Few-Shot Prompting, Large Language Models, In-Context Learning, Code Vulnerability Detection
\end{IEEEkeywords}

\section{Introduction}

Large Language Models (LLMs) have demonstrated strong performance across a wide range of natural language processing tasks, including code understanding and software vulnerability detection \cite{wang2023review}. Fine-tuning these models for specific tasks has become a standard approach to adapting them to specialized domains. However, fine-tuning is resource intensive, may require access to model weights, and entails non-trivial training time and maintenance costs \cite{trad2024prompt,trad2024evaluating}. This can be particularly limiting in security-focused applications, where data distributions evolve frequently and model updates must be efficient and repeatable.

Few-shot prompting has emerged as an alternative that avoids the need for model retraining. In this setup, a small number of labeled input-output examples are embedded directly into the prompt to guide the model during inference. This approach leverages the capabilities of general-purpose LLMs without modifying their internal parameters. While promising, few-shot prompting suffers from high variance depending on the quality and relevance of in-context examples. In complex tasks such as multi-label code vulnerability detection, selecting examples that are semantically mismatched or too generic can lead to degraded performance.

In this work, we explore retrieval-augmented prompting strategies to improve the effectiveness of few-shot learning for code vulnerability detection. Our hypothesis is that providing semantically similar in-context examples can improve prediction accuracy without requiring any model fine-tuning. To evaluate this, we implement and compare three prompting strategies:
\begin{itemize}
    \item \textbf{Random Few-Shot Prompting}: in-context examples are randomly sampled from the training set.
    \item \textbf{Retrieval-Augmented Few-Shot Prompting}: examples are selected based on semantic similarity to the target input using an embedding-based retrieval mechanism.
    \item \textbf{Retrieval-Based Labeling}: labels are inferred by retrieving the most similar examples and propagating their labels directly, without using the LLM for inference.
\end{itemize}

We conducted a detailed evaluation using the Gemini-1.5-Flash model on a multi-label code vulnerability detection dataset. Our results show that retrieval-augmented prompting consistently outperforms both random prompting and retrieval-based labeling. With 20 shots, retrieval-augmented prompting achieves an F1 score of \textbf{74.05\%} and a partial match accuracy of \textbf{83.90\%}, substantially outperforming zero-shot prompting and the other investigated strategies.

To compare with fine-tuning approaches, we consider two settings. First, we fine-tune Gemini-1.5-Flash using Vertex AI on Google Cloud. While this method does not require access to model weights, it incurs monetary cost and offers limited control over training behavior. In this setting, the fine-tuned Gemini model achieves an F1 score of \textbf{59.31\%} and a partial match accuracy of \textbf{53.10\%}, which retrieval-augmented prompting surpasses without any training overhead. 

Second, we fine-tune several smaller open-source models, including DistilBERT, DistilGPT2, and CodeBERT, using full access to weights and local compute resources. Among these, CodeBERT achieves the highest performance (F1: \textbf{91.22\%}, partial match: \textbf{91.30\%}). However, this comes at the cost of training infrastructure, hyperparameter tuning, and ongoing model maintenance. In contrast, retrieval-augmented prompting provides a lightweight and flexible alternative that is easier to deploy and adapt to changing data conditions, when fine-tuning resources are not available.

Our findings demonstrate that semantic retrieval of in-context examples significantly enhances few-shot prompting effectiveness, making it a viable middle ground between zero-shot inference and full model fine-tuning. This study also highlights trade-offs between accuracy, training overhead, and system maintainability, which are important for real-world adoption in secure software development pipelines.

Our main contributions are summarized as follows:
\begin{itemize}
    \item We extend few-shot prompting for code vulnerability detection with a retrieval-augmented approach that selects semantically relevant in-context examples.
    \item We compare this method against random prompting, retrieval-based labeling, zero-shot prompting, and fine-tuning of both large commercial models and small open-source models.
    \item We show that retrieval-augmented prompting achieves substantial gains over other prompting strategies, as well as fine-tuned LLMs like Gemini-1.5-Flash, while requiring no model training.
    \item We present a cost-performance tradeoff analysis that positions retrieval-augmented prompting as a practical alternative to fine-tuning in many real-world settings where resources may not always be available.
\end{itemize}

The rest of the paper is organized as follows: Section~\ref{sec:related_work} reviews related work. Section~\ref{sec:methodology} describes the methodology. Section~\ref{sec:experimental_setup} outlines the experimental setup (dataset, models, and evaluation metrics). Section~\ref{sec:results} presents and analyzes the experimental results. Section~\ref{sec:discussion} discusses the results further and highlights practical implications. Finally, Section~\ref{sec:conclusion} concludes and outlines directions for future research.

\section{Related Work}
\label{sec:related_work}

Machine learning approaches to code vulnerability detection have progressed significantly in recent years, evolving from rule-based static analysis and pattern matching into data-driven models that can capture deeper code semantics. Early work relied on hand-crafted features and syntactic patterns \cite{harer2018automated}, while more recent studies employed neural models such as RNNs, CNNs, and graph neural networks (GNNs), operating over abstract syntax trees (ASTs), control flow graphs (CFGs), or other structured code representations \cite{chakraborty2021deep}. However, these models typically require substantial annotated data and must be retrained when applied to new programming languages, vulnerability types, or domains.

LLMs such as CodeBERT, Codex, and Gemini have demonstrated strong performance on code-related tasks including summarization, generation, and defect detection \cite{sharma2024llms}. Fine-tuning LLMs on task-specific datasets can yield good results, but this process is resource intensive, requires training infrastructure, and complex ongoing maintenance. In hosted settings, fine-tuning via APIs offers a lighter alternative but still incurs monetary cost and limits control over training dynamics and model internals. This is why multiple strategies have been proposed to improve the performance of prompt engineering, including ensemble approaches \cite{trad2024ensemble}, agentic approaches \cite{trad2024large}, and automated prompting techniques \cite{trad2025manual}.

Few-shot prompting has recently emerged as a promising alternative to fine-tuning, enabling LLMs to generalize to new tasks by conditioning on a small number of input–output examples provided in the prompt \cite{logan2021cutting,ma2023fairness}. This approach avoids weight updates and offers greater flexibility. However, the performance of few-shot prompting is highly sensitive to the quality and selection of in-context examples. Previous work showed that random selection of examples can lead to unstable or suboptimal performance, especially in complex or structured prediction tasks \cite{yu2023retrieval}.

Retrieval-augmented generation (RAG) techniques, which dynamically select semantically relevant data using embedding-based retrieval, have shown success in natural language domains such as open-domain question answering and dialogue \cite{siriwardhana2023improving}. However, their application to code-related tasks, particularly structured problems like multi-label vulnerability classification and optimal in-context example selection, remains underexplored.

Our work extends this line of research by conducting a detailed empirical study of retrieval-augmented few-shot prompting strategies in the context of multi-label code vulnerability detection. We benchmark retrieval-augmented prompting against random prompting, direct label transfer via retrieval (retrieval-based labeling), and both zero-shot and fine-tuned LLMs. Importantly, we compare prompting-based approaches not only against proprietary models (e.g., Gemini-1.5-Flash) but also against locally fine-tuned smaller models (e.g., DistilBERT, DistilGPT2, and CodeBERT). Our results demonstrate that retrieval-augmented prompting delivers competitive performance without the need for model retraining, providing a favorable trade-off between cost, flexibility, and predictive accuracy, especially in scenarios where resources for fine-tuning smaller models are unavailable.

\section{Methodology}
\label{sec:methodology}
\begin{figure*}[htbp]
    \centering
    \includegraphics[width=\linewidth]{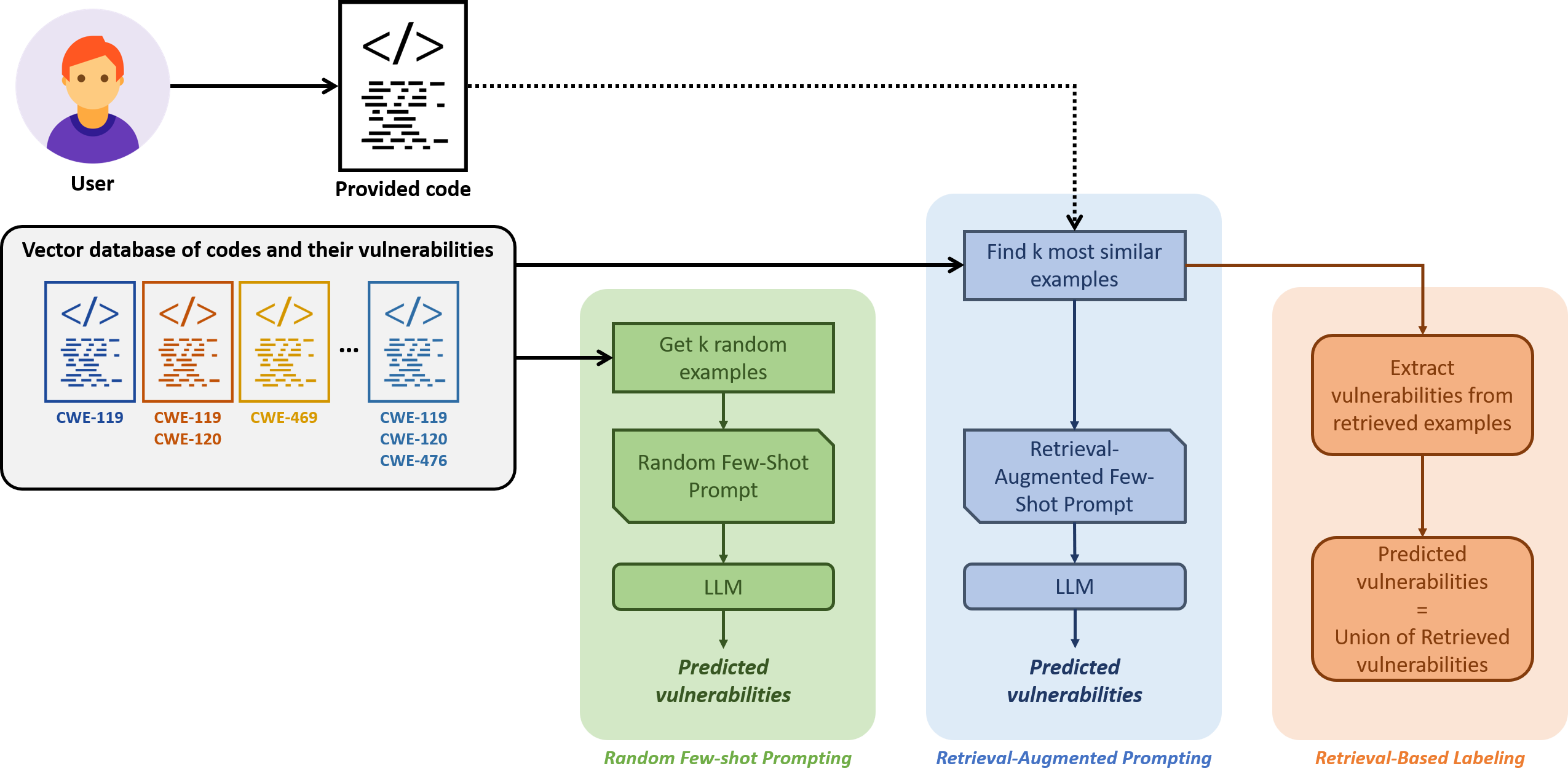}
    \caption{Summary of the three investigated few-shot strategies}
    \label{fig:methodology}
\end{figure*}

This work tackles the task of multi-label code vulnerability detection, where the objective is to identify all relevant vulnerability categories, drawn from a predefined set of Common Weakness Enumerations (CWEs), for a given input code snippet. This setup reflects real-world scenarios where a single code segment may exhibit multiple types of weaknesses. The task is particularly challenging due to the need for deep semantic understanding and precise categorization. Initially, we address this problem with few-shot strategies as shown in Figure \ref{fig:methodology}. 

\begin{enumerate}
    \item \textbf{Random Few-Shot Prompting: } In this approach, a fixed number of in-context examples is randomly sampled from the training pool for each test sample. Each retrieved example consists of a code snippet along with its corresponding vulnerability labels. The selected examples are formatted as input-output pairs and concatenated with the test input to form a single prompt.
    \item \textbf{Retrieval-Augmented Few-Shot Prompting:} 
    This strategy enhances few-shot prompting by selecting in-context examples that are semantically similar to the test input, rather than choosing them at random. 
    Each training example is encoded into a dense vector representation using an embedding model (\texttt{gemini-embedding-001}). 
    We embed each function-level sample as a single vector, ensuring that no code snippet is split across multiple chunks. 
    Both the code snippet and its associated labels are included in the embedding. 
    These vectors are stored in a ChromaDB vector database, which supports efficient cosine similarity search. 
    At inference time, the test input is embedded in the same way, and the top-$k$ most similar training examples are retrieved. 
    We vary $k$ from $1$ to $20$ in order to evaluate how performance scales with the number of retrieved examples. 
    The retrieved examples are formatted as input--output pairs and inserted into the prompt, analogous to standard few-shot prompting. 
    
    \item \textbf{Retrieval-Based Labeling:} 
    In this approach, we bypass the LLM's inference capability altogether. 
    For a given test input, we retrieve the top-$k$ most similar examples from the training pool (again with $k \in [1,20]$) and directly use the union of their labels as the predicted labels. 
    This serves as a baseline for assessing the added value of model inference over simple label propagation.

\end{enumerate}

To provide a more comprehensive evaluation, we also compare the few-shot strategies with zero-shot prompting as well as the fine-tuning approaches. We consider two distinct fine-tuning settings:

\begin{enumerate}
    \item \textbf{Fine-Tuning Gemini-1.5-Flash: } We fine-tune the Gemini-1.5-Flash model using Vertex AI on Google Cloud, which allows supervised training without requiring access to the underlying model weights. The training dataset consists of code snippets paired with their corresponding multi-label vulnerability annotations. The model is fine-tuned using the default hyperparameters set by Google Cloud, which resulted in 10 epochs. To comply with the platform guidelines regarding sample limits, we randomly selected 2,000 examples for the fine-tuning process. While this approach avoids infrastructure setup, it incurs monetary cost and limits control over training parameters. Inference is subsequently performed using the fine-tuned model.

    \item \textbf{Open-Source Model Fine-Tuning: }We fine-tune three smaller open-source transformer models: DistilBERT, DistilGPT2, and CodeBERT. All models are trained on the same dataset using a supervised multi-label classification setup, where the input is a tokenized code snippet and the output is a set of predicted vulnerability types. For consistency, each model is fine-tuned for 10 epochs to enable a fair comparison with the Gemini fine-tuning results. The fine-tuning is performed locally on GPU hardware, providing full control over training configurations such as learning rate, batch size, and evaluation strategy.

\end{enumerate}

This fine-tuning evaluation enables us to assess the performance–cost tradeoff between hosted LLMs, which are convenient but less customizable, and smaller open models, which offer more flexibility at the expense of infrastructure and tuning complexity.

\section{Experimental Setup}
\label{sec:experimental_setup}
\subsection{Dataset}
We use a publicly available dataset of labeled code snippets annotated with software vulnerabilities~\cite{russell2018automated}. Each function-level sample is paired with binary indicators for five possible vulnerability classes: \texttt{CWE-119} (buffer overflow), \texttt{CWE-120} (stack-based buffer overflow), \texttt{CWE-469} (pointer arithmetic error), \texttt{CWE-476} (null pointer dereference), and a catch-all category \texttt{CWE-other}. In addition, many samples are labeled as non-vulnerable. Because functions may contain more than one flaw, the dataset is inherently multi-label. For our study, we focus on the four specific CWE categories with clear definitions and consistent representation: \texttt{CWE-119}, \texttt{CWE-120}, \texttt{CWE-469}, and \texttt{CWE-476}. We exclude non-vulnerable functions to avoid majority-class bias, which would otherwise inflate accuracy scores without reflecting true vulnerability recognition. We also omit \texttt{CWE-other}, since it groups heterogeneous cases and prevents reliable evaluation when a model predicts vulnerabilities not explicitly annotated in the dataset. By restricting attention to well-defined and verifiable categories, we obtain a cleaner evaluation setting that better reflects the model’s ability to capture vulnerability-specific patterns.

\subsection{Models}
We conducted our experiments using both a large proprietary LLM and a set of smaller open-source transformer models. For the proprietary setting, we chose \textit{Gemini-1.5-Flash}, a commercial LLM provided via API by Google. This model offers a favorable balance between cost, speed, and accuracy, making it practical for running extensive controlled experiments while still reflecting the capabilities of modern frontier LLMs. Its efficient inference allows us to explore both zero-shot and few-shot prompting at scale, and we also fine-tune it via Vertex AI using supervised training. Although other LLMs are available, Gemini-1.5-Flash remains a strong and widely deployed baseline, and its stability ensures reproducible results across multiple runs.

For comparison, we also fine-tune three open-source models: DistilBERT, DistilGPT2, and CodeBERT. DistilBERT and DistilGPT2 are compact transformer models pretrained on natural language using masked and autoregressive objectives, respectively \cite{sanh2019distilbert,huggingface-distilgpt2}. CodeBERT is a transformer pretrained jointly on code and natural language, and has been shown to perform well on various software engineering tasks \cite{feng2020codebert}. These models are fine-tuned locally with full access to training configurations, using a single NVIDIA T4 GPU. This setup enables controlled experiments on the vulnerability detection dataset while keeping the results reproducible and comparable to the proprietary LLM setting.

\subsection{Prompt Templates}
\begin{figure}[htbp]
    \centering
    \includegraphics[width=\linewidth]{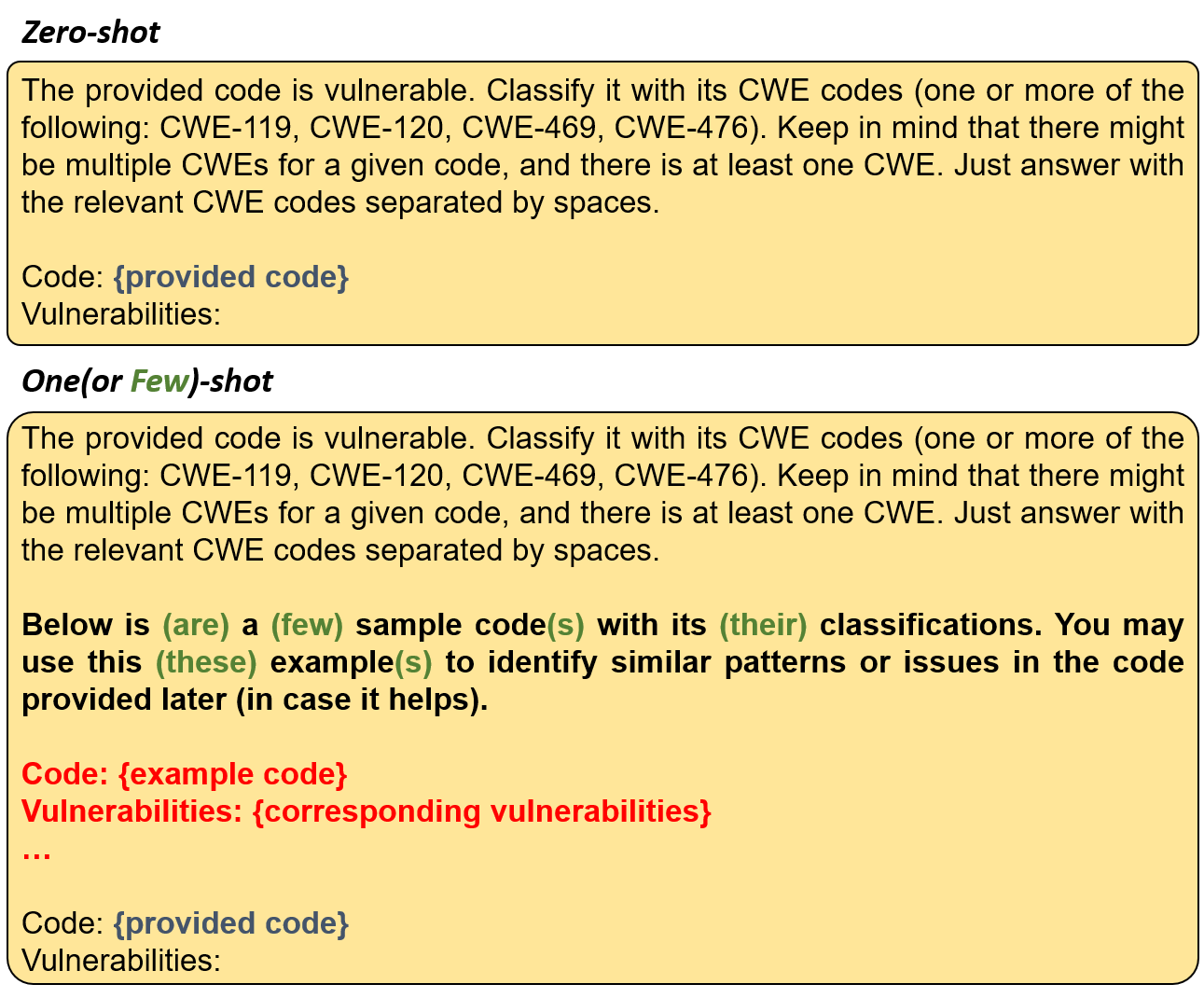}
    \caption{The various prompts used for vulnerability detection}
    \label{fig:prompts}
\end{figure}

We used a consistent prompt template across all few-shot experiments, regardless of how the in-context examples are selected (randomly or via retrieval). Each prompt includes a fixed number of code snippets along with their corresponding vulnerability labels, followed by the test code sample for which the model must generate predictions. To evaluate the impact of providing examples, we also include a zero-shot setting, where the prompt contains only a task description and the test input, with no examples. All prompt formats used in our experiments are illustrated in Figure~\ref{fig:prompts}.

\subsection{Evaluation Metrics}
We evaluated the model predictions using several standard metrics for multi-label classification. These metrics capture different aspects of performance, particularly relevant when each example can be associated with multiple labels simultaneously.

\begin{itemize}
    \item \textbf{Subset Accuracy}: This is a strict metric that measures the proportion of instances where the predicted label set exactly matches the ground-truth label set. A prediction is considered correct only if all correct labels are predicted. This metric is sensitive to partial errors and typically yields lower values in multi-label settings, especially when the number of possible labels is large.

    \item \textbf{Hamming Accuracy}: This is derived from the Hamming loss, which computes the fraction of incorrect labels (both false positives and false negatives) over the total number of labels across all instances. Hamming Accuracy is defined as \(1 - \text{Hamming Loss}\), and it reflects the average correctness of individual label predictions, providing a more forgiving alternative to Subset Accuracy.

    \item \textbf{Partial Match Accuracy}: Also known as label-wise accuracy or overlap accuracy, this metric calculates the proportion of correctly predicted labels per instance, averaged over all instances. It rewards predictions that partially overlap with the ground truth, even if they are not exact matches. This is useful for assessing how often the model is partially correct.

    \item \textbf{Micro Precision, Recall, and F1 Score}: These metrics are computed by aggregating true positives, false positives, and false negatives over all labels and instances. 
    \begin{itemize}
        \item \emph{Precision} measures the proportion of predicted labels that are correct.
        \item \emph{Recall} measures the proportion of true labels that were correctly predicted.
        \item \emph{F1 Score} is the harmonic mean of Micro Precision and Micro Recall.
    \end{itemize}
    These metrics are well-suited for datasets with label imbalance, as they treat each label-instance pair equally.
\end{itemize}

\section{Results}
\label{sec:results}

\begin{figure*}[htbp]
\centering
\begin{subfigure}[b]{0.49\textwidth}
    \centering
    \includegraphics[width=\linewidth]{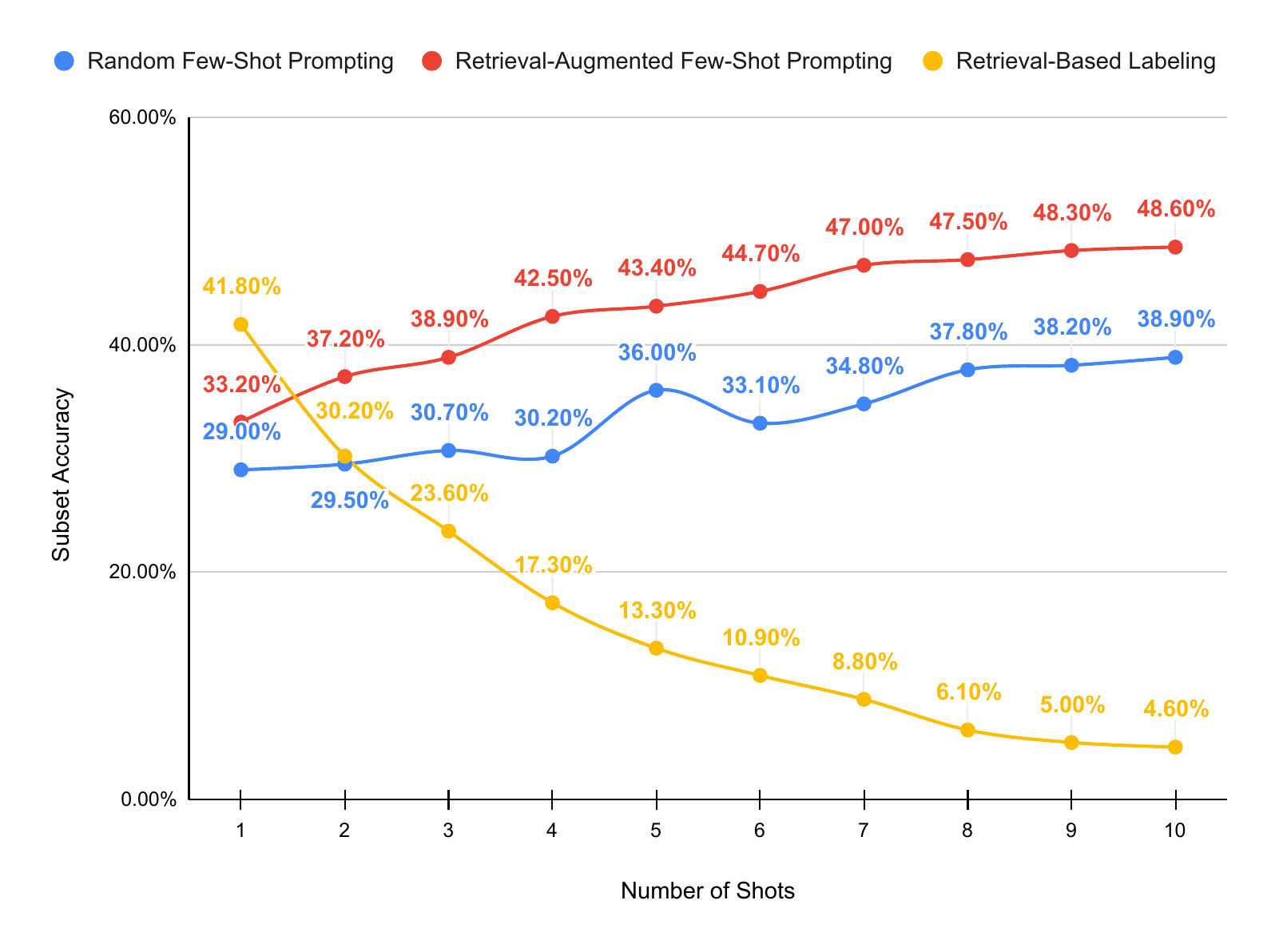}
    \caption{Subset Accuracy}
    \label{fig:subset_accuracy}
\end{subfigure}
\hfill
\begin{subfigure}[b]{0.49\textwidth}
    \centering
    \includegraphics[width=\linewidth]{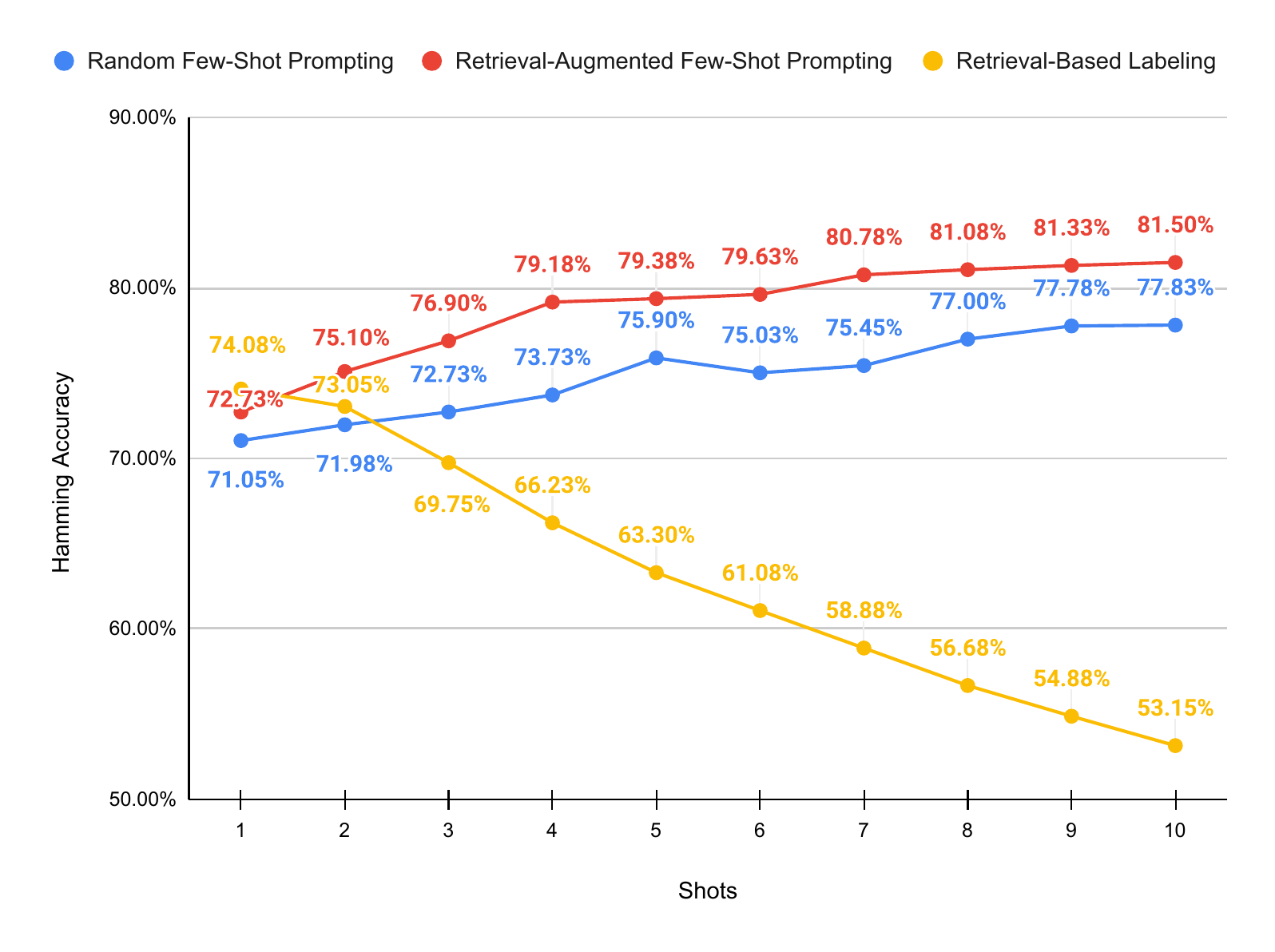}
    \caption{Hamming Accuracy}
    \label{fig:hamming_accuracy}
\end{subfigure}

\vspace{0.5cm} 

\begin{subfigure}[b]{0.49\textwidth}
    \centering
    \includegraphics[width=\linewidth]{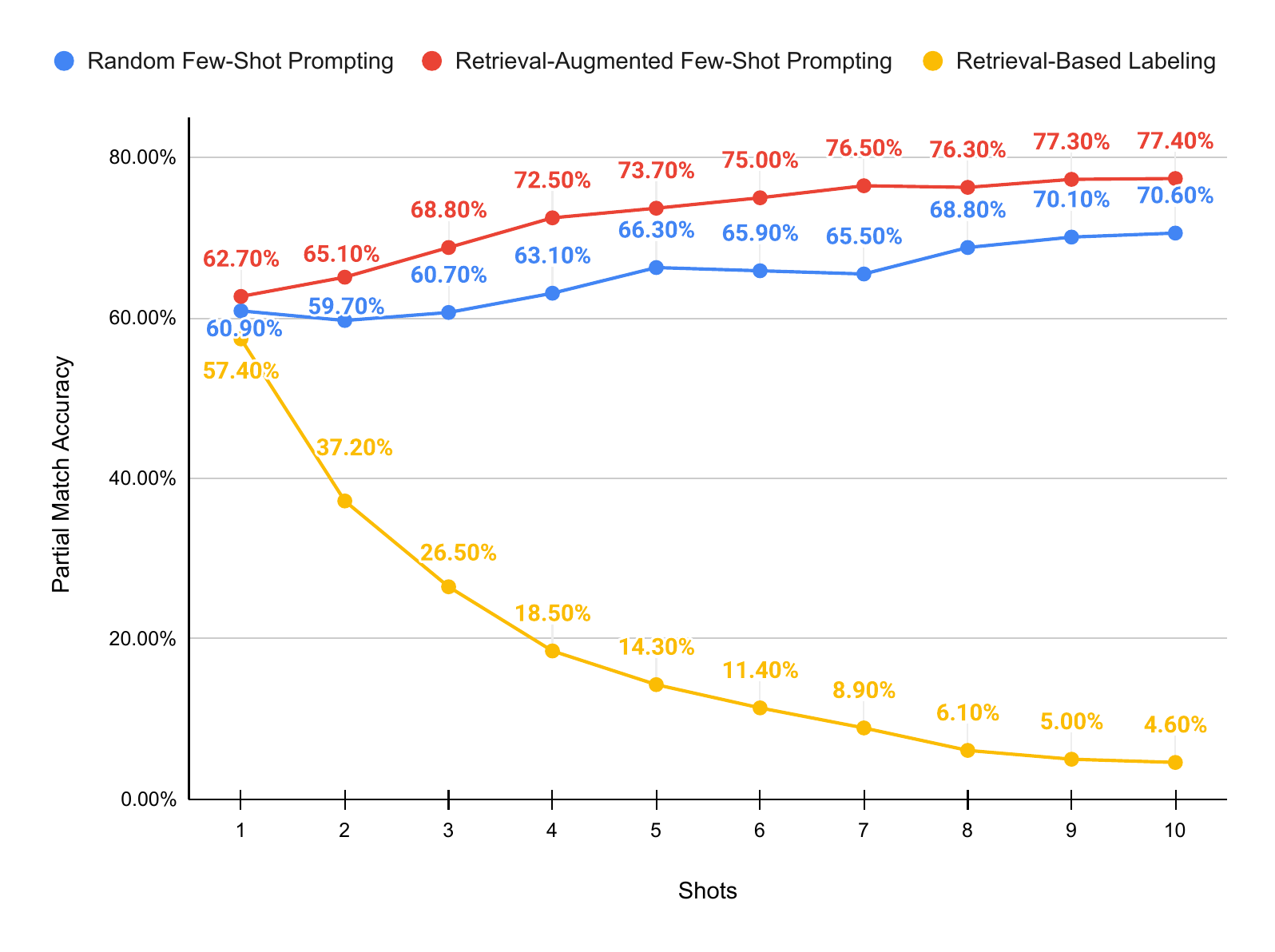}
    \caption{Partial Match Accuracy}
    \label{fig:partial_match_accuracy}
\end{subfigure}
\hfill
\begin{subfigure}[b]{0.49\textwidth}
    \centering
    \includegraphics[width=\linewidth]{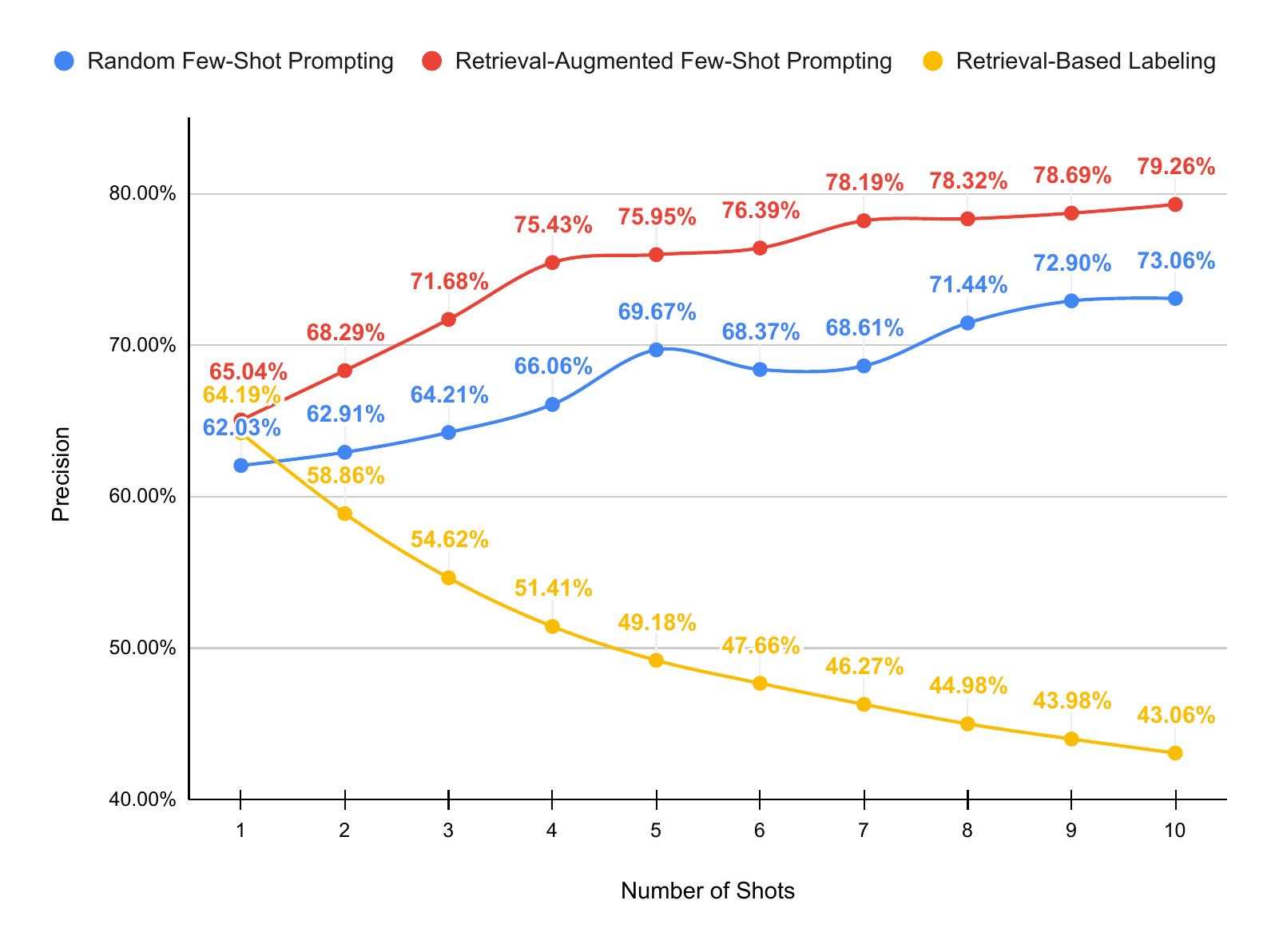}
    \caption{Precision}
    \label{fig:precision}
\end{subfigure}

\vspace{0.5cm} 

\begin{subfigure}[b]{0.49\textwidth}
    \centering
    \includegraphics[width=\linewidth]{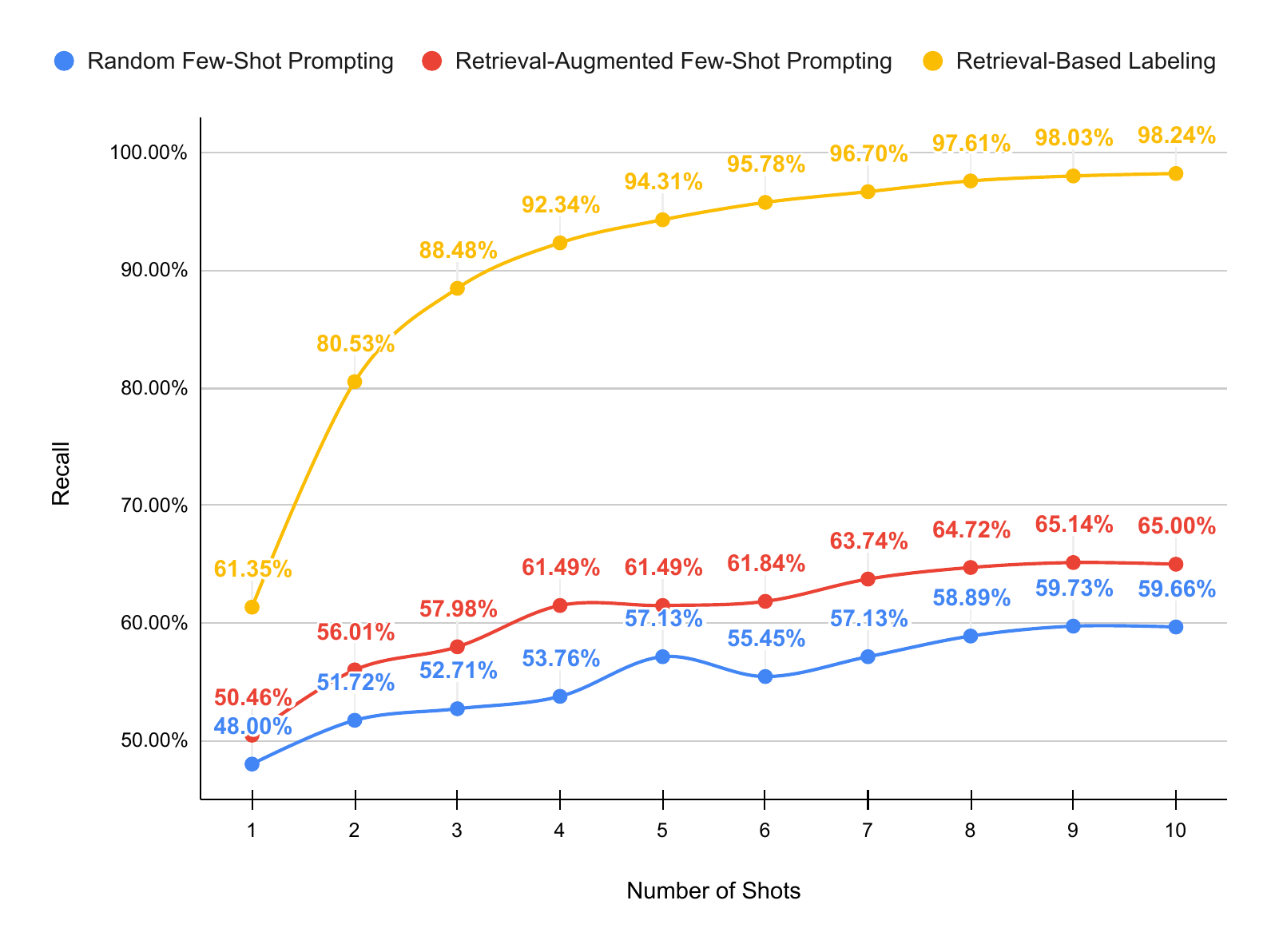}
    \caption{Recall}
    \label{fig:recall}
\end{subfigure}
\hfill
\begin{subfigure}[b]{0.49\textwidth}
    \centering
    \includegraphics[width=\linewidth]{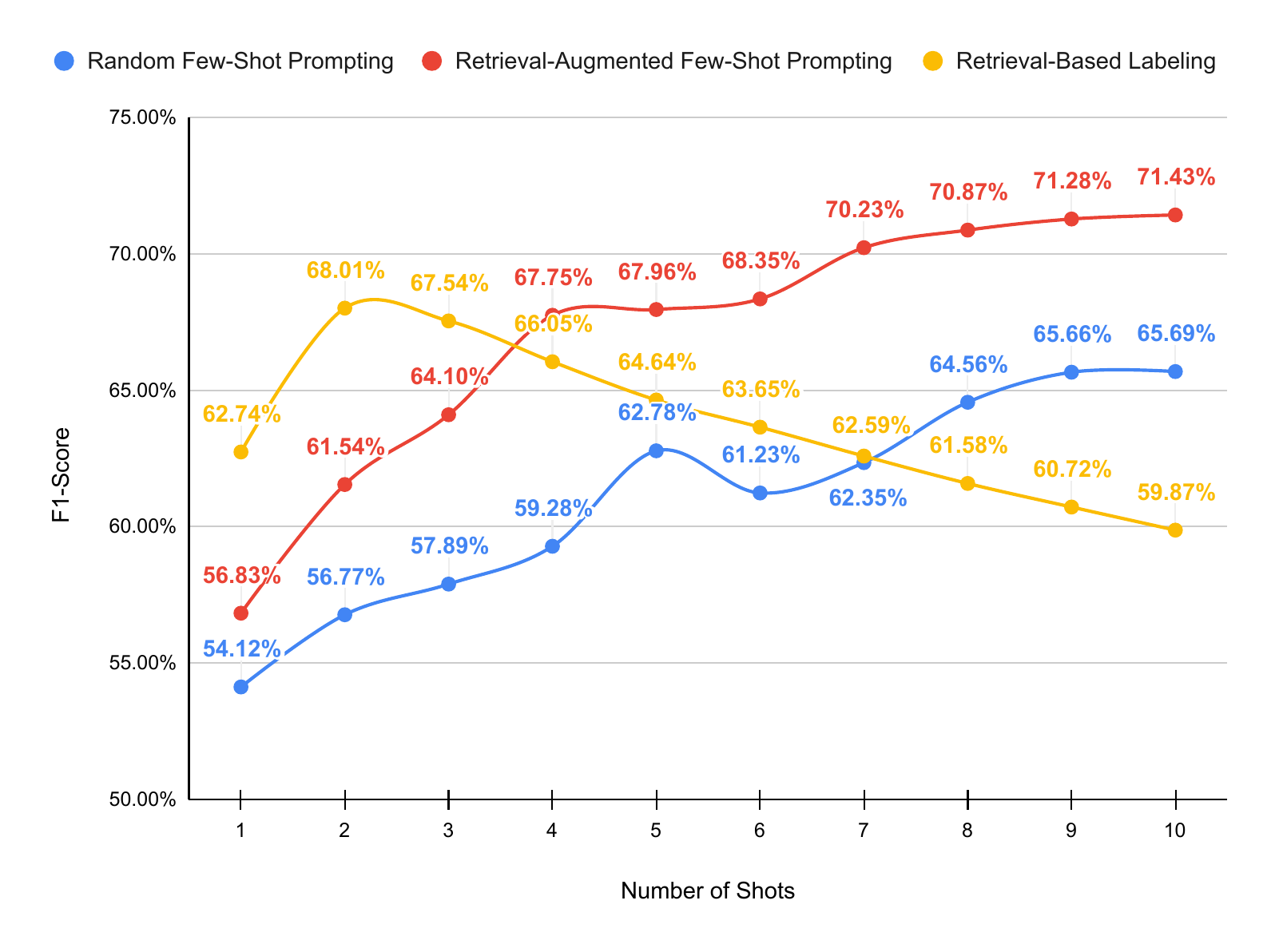}
    \caption{F1 Score}
    \label{fig:f1_score}
\end{subfigure}

\caption{Performance evolution across shot counts (1-10) for the various metrics}
\label{fig:performance_trends}
\end{figure*}

We present a comprehensive evaluation of the different few-shot prompting strategies for vulnerability detection. Figure~\ref{fig:performance_trends} shows how the performance evolves in terms of varying shot counts, from 1 to 10.

\subsection{Overall Performance Trends}

Overall, \textbf{Retrieval-Augmented Few-Shot Prompting} consistently outperforms both random prompting and retrieval-based labeling across all metrics and shot counts. It shows steady and reliable improvements as more semantically relevant examples are added to the prompt, indicating that carefully selected in-context examples provide more informative guidance for the model.

\textbf{Random Few-Shot Prompting} demonstrates moderate gains as the number of shots increases, but its performance remains lower than that of retrieval-augmented prompting at every shot level. This highlights the importance of example relevance in few-shot learning.

\textbf{Retrieval-Based Labeling}, while initially competitive at very low shot counts (particularly at \(k=1\)), exhibits a consistent decline in performance as more examples are used. 

These results demonstrate that retrieval-augmented prompting offers a strong balance between scalability, inference quality, and robustness, and is the most effective strategy among those evaluated. A detailed breakdown of each metric is provided in the following section.

\subsection{Metric-Specific Analysis}

\textbf{Subset Accuracy:} This is the most stringent metric, requires an exact match between the predicted and true label sets. As shown in Figure~\ref{fig:subset_accuracy}, Retrieval-Augmented Prompting demonstrates a consistent upward trend, achieving a maximum of \textbf{48.60\%} at 10 shots—significantly higher than Random Prompting (\textbf{38.90\%}) and Retrieval-Based Labeling (\textbf{4.60\%}). Notably, Retrieval-Based Labeling performs well at \(k=1\) (\textbf{41.80\%}), but its accuracy quickly declines as more examples are added, due to the accumulation of incorrect labels from unrelated samples.

\textbf{Hamming Accuracy:}  This metric evaluates per-label correctness and is more tolerant of partial errors. As shown in Figure~\ref{fig:hamming_accuracy}, Retrieval-Augmented Prompting again leads across all shot counts, reaching \textbf{81.50\%} at 10 shots, compared to \textbf{77.83\%} for Random Prompting and \textbf{53.15\%} for Retrieval-Based Labeling. The gap between retrieval-augmented and random prompting is consistent, confirming the benefits of using semantically similar examples to guide the model more precisely at the label level.

\textbf{Partial Match Accuracy:}  
This metric rewards partially correct predictions by measuring the average overlap between predicted and true label sets. Figure~\ref{fig:partial_match_accuracy} shows that Retrieval-Augmented Prompting achieves the strongest results, increasing steadily to \textbf{77.40\%} at 10 shots, versus \textbf{70.60\%} for Random Prompting. Retrieval-Based Labeling declines sharply beyond \(k=1\), demonstrating that it does not scale well for multiple examples.

\textbf{Precision:}  
This metric reflects the model’s ability to avoid false positives. As shown in Figure \ref{fig:precision}, Retrieval-Augmented Prompting achieves the highest precision at \textbf{79.26\%} (10 shots), indicating that semantically retrieved examples help the model generate more focused and relevant predictions. Random prompting performs reasonably well (\textbf{73.06\%}), while Retrieval-Based Labeling deteriorates rapidly from \textbf{64.19\%} at 1 shot to just \textbf{43.06\%} at 10 shots, as more unrelated labels are aggregated.

\textbf{Recall:}  
Figure \ref{fig:recall} shows that Retrieval-Based Labeling achieves the highest recall (\textbf{98.24\%} at 10 shots) due to over-prediction; it tends to include a superset of correct labels by combining retrieved labels without filtering. However, this comes at the expense of precision and overall quality. Retrieval-Augmented Prompting maintains a balanced recall (\textbf{65.00\%}), outperforming Random Prompting (\textbf{59.66\%}) while avoiding the over-generalization of the label-based baseline.

\textbf{F1 Score: }  
As a harmonic mean of precision and recall, F1 Score offers a comprehensive view of model performance. Figure \ref{fig:f1_score} shows that Retrieval-Augmented Prompting consistently achieves the best F1 scores, reaching \textbf{71.43\%} at 10 shots. Random Prompting follows at \textbf{65.69\%}, while Retrieval-Based Labeling starts strong but falls to \textbf{59.87\%}, unable to maintain performance as more examples are included. The F1 trends further confirm that retrieval-augmented prompting offers the best balance between label coverage and accuracy.

\subsection{Comparison to Zero-shot Prompting and Fine-Tuning}
Table~\ref{tab:results} presents a comparative analysis of the retrieval-augmented prompting strategy against zero-shot prompting and several fine-tuning baselines, including both proprietary and open-source models. This comparison highlights not only differences in predictive performance but also important considerations such as training time, cost, and resource requirements.

\begin{table*}[htbp]
\centering
\caption{Performance comparison at 10-shot configuration and beyond}
\label{tab:results}
\resizebox{2\columnwidth}{!}{%
\begin{tabular}{lcccccccccc}
\toprule
\textbf{Method} & \textbf{Subset Acc.} & \textbf{Hamm. Acc.} & \textbf{Partial Match} & \textbf{Precision} & \textbf{Recall} & \textbf{F1} & \textbf{Train time} & \textbf{Train cost} & \textbf{Resources} \\
\midrule
Zero-shot Prompt & 17.40\% & 52.38\% & 20.30\% & 34.65\% & 38.23\% & 36.35\% & 0 & 0 & None \\
Retrieval-Aug. 10-shot & 48.60\% & 81.50\% & 77.40\% & 79.26\% & 65.00\% & 71.43\% & 0 & 0 & None \\
Retrieval-Aug. 20-shot & 52.60\% & 83.60\% & 83.90\% & 84.71\% & 65.78\% & 74.05\% & 0 & 0 & None \\
Fine-tuned Gemini-1.5-Flash & 33.10\% & 71.53\% & 53.10\% & 60.32\% & 58.33\% & 59.31\% & 47m 51s & \$36.61 & Cloud Sub. \\
DistilBERT & 69.30\% & 89.08\% & 89.70\% & 91.15\% & 76.74\% & 83.33\% & 20m 13s & 0 (if local) & GPU \\
DistilGPT2 & 80.70\% & 92.75\% & 89.80\% & 92.06\% & 87.14\% & 89.53\% & 24m 33s & 0 (if local) & GPU \\
CodeBERT & 84.80\% & 93.88\% & 91.30\% & 93.12\% & 89.39\% & 91.22\% & 40m 07s & 0 (if local) & GPU \\
\bottomrule
\end{tabular}%
}
\end{table*}

As expected, zero-shot prompting yields the weakest performance across all metrics. Without any in-context examples, the model struggles to generalize to the vulnerability detection task, achieving only \textbf{36.35\%} F1 score and \textbf{20.30\%} partial match accuracy. These results underscore the importance of task-specific context, especially in complex structured prediction settings such as multi-label vulnerability classification.

Retrieval-augmented prompting significantly outperforms the zero-shot baseline. At 10 shots, it achieves an F1 score of \textbf{71.43\%}, more than doubling the zero-shot performance, and a partial match accuracy of \textbf{77.40\%}. At 20 shots, performance improves further, reaching \textbf{74.05\%} F1 and \textbf{83.90\%} partial match accuracy. Importantly, these results are obtained \textit{without any training time or cost}, using only inference via prompt construction. This highlights the method’s strong cost-efficiency and ease of deployment, particularly in scenarios where model weights are inaccessible or retraining is impractical.

Fine-tuning Gemini-1.5-Flash using Vertex AI, which allows task-specific adaptation without access to internal weights, is convenient and does not require coding expertise. While fine-tuning improves performance relative to zero-shot prompting (F1: \textbf{59.31\%}), it remains substantially below the retrieval-augmented prompting results (F1: \textbf{74.05\%}). Additionally, the fine-tuning process incurred \textbf{47 minutes of training time} and a cost of \textbf{\$36.61}, requiring a paid cloud subscription. These findings suggest that, while fine-tuning commercial LLMs is a viable option, it is less efficient and less effective than high-quality prompting with example retrieval.

On the other hand, fine-tuned open-source models outperformed all prompting strategies, but at the cost of additional infrastructure and training effort. Among these models, CodeBERT achieved the highest overall performance, with an F1 score of \textbf{91.22\%} and a partial match accuracy of \textbf{91.30\%}. Fine-tuning was conducted locally using GPU resources, which incurs no direct monetary cost but does require access to compute infrastructure and tuning expertise. In scenarios where such resources are not available locally, using cloud-based alternatives can introduce substantial computational costs.

\section{Discussion}
\label{sec:discussion}

The results presented in this work demonstrate that retrieval-augmented prompting is a highly effective strategy for multi-label code vulnerability detection. By incorporating semantically relevant examples into the prompt, we can substantially improve model performance without any fine-tuning, retraining, or infrastructure overhead. This finding has important implications for both the research community and practitioners deploying LLMs in software engineering pipelines.

\subsection{Effectiveness of Example Selection}

The performance differences between the three prompting strategies highlight the critical role of example quality in few-shot learning. Retrieval-based labeling performs well at very low shot counts (\(k=1\) or \(k=2\)), likely due to high similarity between the query and retrieved examples. However, as more examples are included, this method quickly degrades. Since it bypasses model inference and aggregates labels directly, it tends to over-predict and include irrelevant classes, resulting in declining precision and overall F1 score.

Random few-shot prompting provides a more stable alternative, offering clear gains over zero-shot prompting. However, its lack of semantic alignment with the test input limits its effectiveness. Our results show that retrieval-augmented prompting consistently outperforms random prompting across all shot counts and metrics. The advantage is most pronounced starting at moderate shot counts (e.g., 3–5 shots), where even a small number of relevant examples leads to substantial improvements.

\subsection{Retrieval-Augmented Prompting vs. Fine-Tuning}

Perhaps one of the most impactful findings of this study is that retrieval-augmented prompting can outperform fine-tuned LLMs, including proprietary models such as Gemini-1.5-Flash, even at low shot counts. For example, retrieval-augmented prompting surpasses the fine-tuned Gemini model at just 2 shots (F1: \textbf{61.54\%} vs \textbf{59.31\%}), while random prompting requires 5 shots to reach the same level. This highlights the parameter-efficiency and scalability of example-based prompting, particularly when augmented with semantic retrieval.

While fine-tuned open-source models like CodeBERT still achieve the highest absolute performance, they require significant setup: access to GPUs, labeled training data, tuning expertise, and time. In contrast, retrieval-augmented prompting requires none of these. It offers near plug-and-play capabilities with zero cost and zero training time, making it especially suitable for environments where latency, resources, or engineering time are limited.

\subsection{Practical Implications}

From a deployment perspective, these findings offer a clear methodology selection framework:

\begin{itemize}
    \item \textbf{For low-resource or real-time applications}, retrieval-augmented prompting offers a practical alternative to fine-tuning. It delivers competitive performance without infrastructure requirements.
    
   \item \textbf{In scenarios with limited labeled examples}, such as new projects or unseen code domains, retrieval-augmented prompting scales more effectively than retrieval-based labeling and random prompting. In such cases, fine-tuning may not be feasible due to insufficient data or infrastructure constraints.

    \item \textbf{When maximizing accuracy is critical and sufficient resources are available}, fine-tuning domain-specific models like CodeBERT remains the most effective approach. However, this comes with the added burden of ongoing maintenance, retraining, and infrastructure support.

\end{itemize}

\section{Conclusion and Future Work}
\label{sec:conclusion}

In this work, we investigated the effectiveness of retrieval-augmented prompting for multi-label code vulnerability detection using LLMs. We compared three prompting strategies (random few-shot prompting, retrieval-based labeling, and retrieval-augmented prompting) over six evaluation metrics. Our results show that retrieval-augmented prompting consistently outperforms both random prompting and zero-shot baselines, and even surpasses fine-tuned proprietary LLMs such as Gemini-1.5-Flash, despite requiring no model retraining or additional cost.

In addition, we benchmark our approach against fine-tuned open-source models, including DistilBERT, DistilGPT2, and CodeBERT. While these models achieved higher absolute performance, they required dedicated training infrastructure and careful tuning. In contrast, retrieval-augmented prompting offers a competitive and resource-efficient alternative that is well-suited for practical deployment, especially in low-resource or latency-sensitive environments.

Our findings highlight the importance of semantically relevant example selection in few-shot prompting and position retrieval-augmented prompting as a robust method for structured code analysis tasks. The method is simple to implement, scalable, and does not rely on access to model internals, making it highly adaptable across domains and platforms. That said, when compute resources are readily available and latency is less of a concern, fine-tuning smaller, open-source models such as CodeBERT can be advantageous, particularly in production environments where consistent, high-accuracy predictions are required, and model updates can be managed over time.

Several directions remain open for exploration. First, we plan to investigate more advanced retrieval mechanisms to further improve the quality of selected examples. Second, we aim to extend this framework to other structured prediction tasks in software engineering, such as API misuse detection, code summarization, and defect localization. These future extensions will help generalize and strengthen the utility of retrieval-based prompting as a lightweight and effective alternative to traditional fine-tuning in real-world code analysis pipelines.

\section*{Acknowledgment}

The authors would like to acknowledge that this work has been supported by the Maroun Semaan Faculty of Engineering and Architecture (MSFEA) at the American University of Beirut (AUB).

\bibliographystyle{ieeetr}
\bibliography{biblio}

\end{document}